\documentclass{JAC2000}
\usepackage{epsf}
\usepackage{daniel}
\title{SIMULATION OF AN INTRA-PULSE INTERACTION POINT FEEDBACK FOR FUTURE LINEAR COLLIDERS}
\author{D. Schulte, CERN, 1211 Geneva, Switzerland}
\begin{document}
\maketitle
\begin{abstract}
In future normal-conducting linear colliders,
the beams will be delivered in short bursts
with a length of the order of $100\u{ns}$. The pulses will be
separated by several ms. In order to maintain high
luminosity, feedback is necessary on a pulse-to-pulse basis.
In addition, intra-pulse feedback that can correct beam
positions
and angles
within one pulse seem technically feasible.
The likely performances of different feedback options are simulated for
the NLC (Next Linear Collider~\cite{c:nlc}) and CLIC (Compact Linear
Collider~\cite{c:clic}).
\end{abstract}

\section{Introduction}
A vertical position displacement between the beam centres at the interaction
point (IP) will cause luminosity reduction.
Two main sources of beam jitter at the interaction point IP are expected.
Firstly, the beam entering the
final focus system may jitter in angle and position. At the IP, the
resulting vertical position error,
normalised to the beam size, and the resulting angle error,
normalised to the beam divergence, are expected to have the same size.
Secondly, transverse jitters of the final focus magnets, especially of the
final doublet, will mainly change the position of the beams
at the IP, not so much the angle. The jitter at the IP can
thus be described by
\begin{displaymath}
\left(\frac{\langle(\Delta y)^2\rangle}{\sigma^2_y}\right)=
\left(\frac{\langle(\Delta{y^\prime})^2\rangle}{\sigma^2_{y^\prime}}\right)
+\left(\frac{\langle(\Delta_{ffs}{y})^2\rangle}{\sigma^2_{y}}\right).
\end{displaymath}
Here, $\Delta y$ and $\Delta y^\prime$ are the offset and angle error of the
beam at the IP, $\sigma_y$ and $\sigma_{y^\prime}$ are beam
size and divergence, also at the IP.
$\Delta_{ffs}y$ is the contribution to the position error
due to the final focus system. If it is large, the effect of the angle at the
IP can be neglected.

\section{Beam-Beam Interaction}
When the beams collide with a vertical offset, they will receive a strong kick
from the beam-beam interaction. The angle of the outgoing beam can therefore be
used to measure the relative positions of the beams.
The dependence of kick angle and luminosity on the position and initial angle
have been simulated with the program {\sc Guinea-Pig}~\cite{c:gp}, varying
both parameters.
\begin{figure}
\epsfxsize=8cm
\centerline{\epsfbox{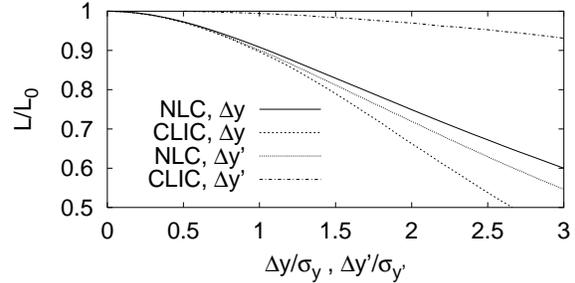}}
\caption{The luminosity as a function of the beam offset and
angle at the IP. CLIC is not very sensitive to $\Delta y^\prime$ because the
vertical beta-function at the IP is much larger than the bunch length.}
\label{f:lumi}
\end{figure}
\begin{figure}
\epsfxsize=8cm
\centerline{\epsfbox{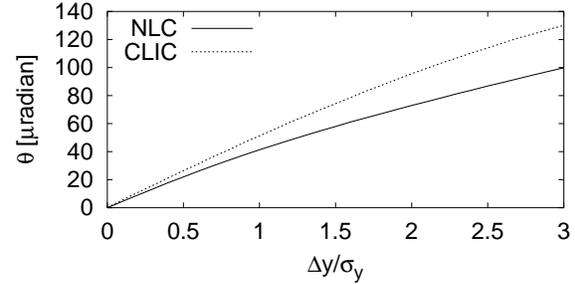}}
\caption{The kick angle $\theta$ as a function of the beam offset.}
\label{f:angle0}
\end{figure}
The luminosity $L$ as a fraction of the nominal $L_0$, is shown in
Fig.~\ref{f:lumi},
as a function on the relative beam position error and beam angle error.
The kick angle is shown in Fig.~\ref{f:angle0} as a function of the offset.
If the beams collide without an offset but with an angle, their initial angle
is roughly preserved in the beam-beam interaction.
For comparison: the beam divergence is
$\sigma_{y^\prime}\approx26\u{\mu radian}$ for NLC and
$\sigma_{y^\prime}\approx11.7\u{\mu radian}$ for CLIC.

\section{Position Feedback Model}
\begin{figure}
\centerline{
\epsfxsize=6.5cm
\epsfbox{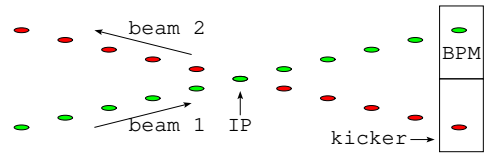}}
\caption{View of the feedback system from above. The beams collide with a
fixed horizontal angle $\theta_c$.
The BPM measures the vertical position of beam 1 and the
kicker corrects beam 2 accordingly.
}
\label{f:scheme}
\end{figure}
In order to have a fast correction, corrector and beam-position
monitor (BPM) need to be located close together. Here, they are located
on the same side of the IP at a distance of
$1.5\u{m}$, see Fig.~\ref{f:scheme}.
Thus the correction is not applied to the measured beam
but to the other one. This significantly reduces the time necessary to
transport the signal from the BPM to the kicker. The
feedback response time $\tau_d$ is given by
\begin{equation}
\tau_d=\tau_p+\tau_k+\tau_{pf}+\tau_{kf}+\tau_s
\label{e:latency}
\end{equation}
Here, $\tau_p$ is the time the BPM electronics needs to measure the beam
offsets and to process the data, $\tau_k$ is the response time of the kicker
and $\tau_s$ is the transport time of the signal from BPM to kicker.
$\tau_{pf}$ and $\tau_{kf}$ are the times of flight from the IP
to the BPM and from the kicker to the IP, respectively.
In the following, a total of $\tau_d=20\u{ns}$ is assumed, half of which is
due to $\tau_{pf}+\tau_{kf}$. The pulse lengths are $100\u{ns}$ in CLIC and
$266\u{ns}$ in NLC.

The hardware for this feedback has not yet been designed. With a solid state
amplifier it should be possible to correct an offset of
$2\sigma_y$~\cite{c:gunter}, with an additional stage of tube amplification
this may even be extended to $20\sigma_y$~\cite{c:joe}.

It is assumed that the feedback changes the beam position by $\delta y$ after
each measured bunch according to
\begin{displaymath}
\delta y=g\frac{\theta}{\sigma_{y^\prime}}\sigma_y
\end{displaymath}
for a measured angle $\theta$.
The gain factor $g$ is chosen to give optimal performance.
The additional crossing angle, that results from the correction
is orders of magnitude smaller than the beam divergence and can be neglected.

\section{Results of Position Feedback}
\begin{figure}
\epsfxsize=8cm
\centerline{\epsfbox{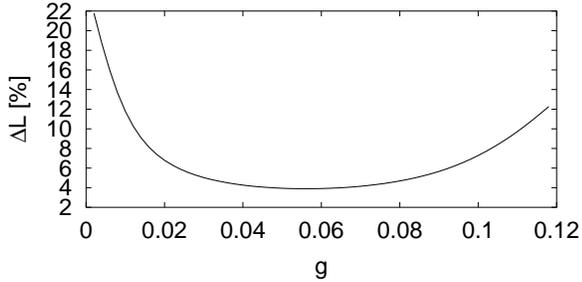}}
\caption{The luminosity loss in NLC (with feedback) for a beam position error
as a function of the gain $g$.}
\label{f:gain}
\end{figure}

Here, only position errors are considered.
First NLC is discussed.
In Fig.~\ref{f:gain}, the luminosity loss with a beam offset
$\Delta y=2\sigma_y$ is shown as a function of the gain $g$.
As can be seen, $g=0.06$ seems a good
choice. Very small gains lead to a slow correction, very large ones to an
over-correction. Both result in a larger luminosity loss.
With $g=0.06$, the luminosity loss is reduced by a factor 6, compared
to the case without feedback. For
a smaller offset of $\Delta y=1/8\sigma_y$ about the same factor is found.

Two main sources of noise can lead to an increased luminosity loss with
feedback: a bunch-to-bunch position jitter of the incoming beam, and
the position resolution of the BPM. For the chosen gain $g=0.06$,
the additional loss induced by the feedback is very small, compared to
the case without feedback.
To estimate the required BPM resolution, simulations are performed with
perfect beams and a position error of the BPM of $\sigma_{BPM}=15\u{\mu m}$
for a single bunch. The luminosity loss, averaged over 100 cases, is only
$\Delta L/L=0.7\times10^{-3}$.
The limit on the BPM resolution seems therefore not to be very stringent
compared to the resolutions that must be achieved in other parts of the
machine.

For a very large offset of $\Delta y=12\sigma_y$, the luminosity
without feedback, is only
$3.5\%$ of the nominal value. If the feedback has the required
correction range, it can recover $73\%$ of the full luminosity. For the
experiment, this can make the difference between a complete failure and still
acceptable running conditions.

For CLIC, the machine with a centre-of-mass energy of $1\u{TeV}$ is
simulated.
At higher energies, $E_{cm}=3\u{TeV}$ or $E_{cm}=5\u{TeV}$,
a large number of electrons and positrons will be
produced during the collision of the two beams, in a process
called coherent pair creation~\cite{c:back};
already at $E_{cm}=3\u{TeV}$, the number of these particles
is about $20\u{\%}$ of the number of beam particles.
They induce a strong signal in the BPM, and due to their large angle
could even hit it.
Their properties need to be studied in detail before one can suggest a feedback
for the high energy machines.

In CLIC at $E_{cm}=1\u{TeV}$,
the feedback response time is assumed to be the same as in NLC.
With the optimum gain $g=0.005$, the luminosity loss is reduced by a factor 3.
This is not as good as in NLC, since the bunch trains are shorter in CLIC.
A BPM resolution of $\sigma_{BPM}=15\u{\mu m}$
leads to a luminosity loss of only $\Delta L/L=1.2\times10^{-4}$.
This is better than in NLC because of the lower
gain and the slightly larger kick angle for an offset of
$\Delta y=\sigma_y$.

\section{Influence of Angles}
If the beams at the IP have angle jitters, this reduces the
luminosity. In addition, the BPM measures the additional angle and the feedback
tries to correct a non-existing offset.
The latter problem can be solved by measuring the incoming beam angle
error and subtracting it from the value measured by the feedback.
Two options are discussed in reference~\cite{c:feedback}, one suggested by
M.~Breidenbach. Both have some difficulties and neither
correct the angle error, but only its effect on the position feedback.
As shown below, this is not sufficient, because the luminosity loss will
stay large. If the angle jitter is significant,
an additional angle feedback is needed for each beam, as described below.

\section{Angle Feedback Model}
\begin{figure}
\epsfxsize=8cm
\epsfbox{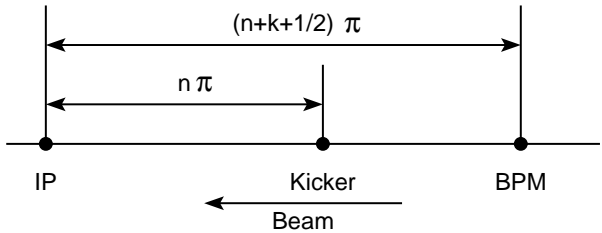}
\caption{Schematic layout of the angle feedback.}
\label{f:angle}
\end{figure}
Each angle feedback consists of a BPM and a strip-line kicker which are placed
in the beam delivery section before the detector, see Fig.~\ref{f:angle}.
This assumes that the angle
jitter is created before this system, as is to be expected.
The BPM has to be at a phase advance of $(n+k+\frac{1}{2})\pi$ from the IP,
where an angle error at the IP can be measured as a position error.
The kicker has to be closer to the IP, at $n\pi$,
to be able to transport the signal in the same direction as the beam.
Here, the angle at the IP can be corrected by applying a kick.
One needs large beta-functions, at the BPM to have a good signal, and
at the kicker to have a smaller divergence and thus correction angle.
Possible positions exist in the beam delivery system~\cite{c:tor}.
The kick angles have to be significantly larger than for the offset
feedback~\cite{c:feedback}, and it may be difficult to achieve this.

This feedback is relatively simple, and uses a constant gain for each bunch.
The response time $\tau_d$ is given by equation~(\ref{e:latency}).
In the present case $\tau_{pf}$ is negative,
since the beam reaches the BPM before the IP. With signal
transmission at the speed of light, one would obtain
$\tau_s+\tau_{pf}+\tau_{kf}=0$ and consequently $\tau_d=\tau_p+\tau_k$.
In the following, $\tau_d=15\u{ns}$ is assumed.

\section{Results of Angle Feedback}
\begin{figure}
\epsfxsize=8cm
\epsfbox{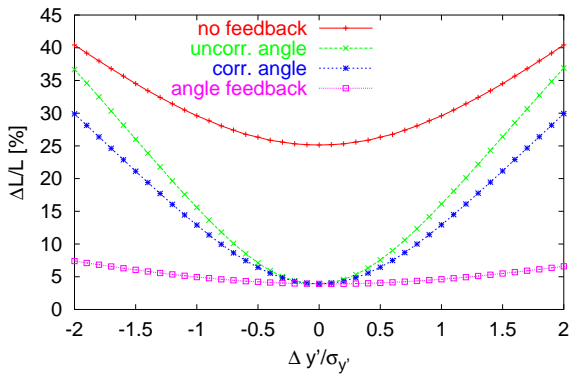}
\caption{The total luminosity loss as a function of the initial angle of the
measured beam. The beam-beam position separation in the interaction point
is $\Delta_y=2\sigma_y^*$.}
\label{f:fig1}
\end{figure}

The angle feedback is simulated for NLC.
The optimum gain is determined in the same way as
for the position feedback. If only angle errors were present, the luminosity
loss would be reduced by a factor 6, as for the position feedback.

The required resolution for the BPM depends on the vertical beta-function
at its position. It must correspond to a resolution of the beam angle
in the IP of $0.2\sigma_{y^\prime}$, to achieve a luminosity loss
of only $\Delta L/L=10^{-3}$ for perfect beams.

Finally, the combination of angle and position error is considered.
Figure~\ref{f:fig1} shows the fractional luminosity loss for a constant beam
position
error of $\Delta y=2\sigma_y$ as a function of the angle error.
If no feedback is used, the luminosity loss is high. An additional angle error
can increase it even more.
If only a position feedback is used, which does not correct the
angle error of the incoming beam, the luminosity loss is small as long as
the angle errors are small. If $\Delta y^\prime/\sigma_{y^\prime}$ becomes
comparable to $\Delta y/\sigma_y$, the loss is almost the same as without
feedback.
If one measures the incoming angle, and subtracts it from the measured value,
the situation does not improve very much.
If finally,
a position feedback at the IP and an angle feedback for each beam
are used, the luminosity loss is significantly reduced,
independent of the initial angle error.

\section{Conclusion}
If the appropriate hardware can be built, the intra-pulse feedback at the
interaction point offers a reduction of the luminosity loss due to
pulse-to-pulse jitter by a factor of about 6 in NLC and 3 in CLIC.
Even in case of a very large offset of 12 times the beam size,
more than $70\u{\%}$ of the luminosity is recovered in NLC. Without
feedback the luminosity would be almost zero.

If the angle jitter is significant, it is not sufficient to correct the
measured kick angle accordingly. To reduce the luminosity loss due to the
angle errors, the described angle feedback is necessary.
Whether it is feasible needs to be studied.

\section{Acknowledgements}
I would like to thank T. Raubenheimer for inviting me to SLAC, where I did
most of the work presented. I am grateful to M.~Breidenbach, P.~Emma,
J.~Frisch, G.~Haller, T.~Raubenheimer and P.~Tennenbaum for very helpful
discussions.

\end{document}